\documentclass{aastex}
\input{epsf.sty}
\usepackage{emulateapj5}
\def\solar{\ifmmode_{\mathord\odot}\else$_{\mathord\odot}$\fi} 
\def\kms{km\thinspace s$^{-1}$}     
\def\deg{\ifmmode^\circ\else$^\circ$\fi}  
\def\arcs{\ifmmode {'' }\else $'' $\fi}  
\def\arcm{\ifmmode {' }\else $' $\fi}    
\def\mstar{M$_{HI_\ast}$}
\def\lya{Ly-$\alpha$} 
\def\msolar{M$_\odot$}

\received{2002 February 4}
\begin{document}

\title{The Contribution of HI-rich Galaxies to the Damped \lya\ Absorber 
Population at $z=0$}

\author{Jessica L. Rosenberg \& Stephen E. Schneider}
\affil{Center for Astrophysics \& Space Astronomy, Department of Astrophysical
and Planetary Sciences, University of Colorado, Boulder, CO 80309}
\affil{Department of Astronomy, University of Massachusetts, Amherst, MA 01003}

\begin{abstract}
We present a study of the expected properties of the low redshift damped
\lya\  absorber population determined from a sample of HI-selected
galaxies in the  local universe. Because of a tight correlation between
the HI mass and HI cross-section, which we demonstrate spans all galaxy
types, we can use our HI-selected  sample to predict the properties of the
absorption line systems. We use measurements of the number density and HI
cross-section of galaxies to show that the total HI cross-section at
column  densities sufficient to produce damped \lya \ absorption is
consistent with no  evolution of the absorber population. We also find
that the $dN/dz$ distribution  is dominated by galaxies with HI masses
near 10$^9$ \msolar. However, because  of the large dispersion in the
correlation between HI mass and stellar luminosity, we find that the
distribution of $dN/dz$ as a function of L$_J$ is  fairly flat.
Additionally, we examine the line widths of the HI-selected galaxies and show
that there may be evolution in the kinematics of HI-rich galaxies, but
it is not necessary for the higher redshift population to contain a greater
proportion of high mass galaxies than we find locally.
\end{abstract}

\section{Introduction}

Damped Lyman-$\alpha$ absorption-line systems (DLAs) seen against
background quasars are the highest column density absorbers, N$_{HI} > 2
\times 10^{20}$ cm$^{-2}$, and are thought to be associated with galaxies.
Often they are assumed to be the disks of large spiral galaxies or their
progenitors (Wolfe 1995), but there is growing evidence that not all DLAs
are associated with bright spirals (e.g. Cohen 2001, Lanzetta et al. 1997,
Miller et al. 1999, Colbert \& Malkan 2001). The difficulty in studying
the properties of these systems arises, in large part, because there are
not many systems known at  low-$z$: only 9 DLAs have been found with $z <
0.5$  (Rao \& Turnshek 2000, Bowen et al. 2001a, Steidel et al. 1994,
Lanzetta et al. 1997, and Le Brun et al. 1997), and only 2 of those are at
$z<0.1$ where  detailed observations are possible. Absorption line studies
do not provide adequate statistical samples at low-$z$ because their
pencil beams survey only small volumes of local space.

Many of the identified DLA candidates have low luminosity or low surface
brightness, so it has been suggested that the absorption line population
may be biased against detecting bright spiral galaxies. For example,
bright spiral  galaxies might contain enough dust to obscure the
background QSO (Fall \& Pei 1993). However, no low redshift comparison
sample exists to provide an indication of what we should expect to see. In
this study we construct a local comparison sample from a blind 21 cm
survey we carried out at the Arecibo  Observatory\footnote{The Arecibo
Observatory is part of the National Astronomy  and Ionosphere Center,
which is operated by Cornell University under cooperative  agreement with
the National Science Foundation.} (Rosenberg \& Schneider 2001)  and use
it to predict the characteristics of the low redshift DLA population. 

Several previous attempts have been made to predict the properties of DLAs
in the local universe based on known properties of nearby galaxies. Rao \&
Briggs (1993) and Rao, Turnshek \& Briggs (1995) combined optical
luminosity functions for different morphological types with a linear 
relationship between HI mass and optical luminosity. They found that
nearly all of the  galaxy cross-section comes from large, bright spirals
and that the HI  cross-section implied by these galaxies would indicate a
substantial decline in total HI cross-section since $z$=0.5.  Salucci \&
Persic (1999) also attempted to account for the local population of DLAs
by combining HI observations with optical luminosity functions. They found
that large spirals contain most of the HI mass density in the local
universe and that their mass density is similar to what is found in the
damped absorbers. In parallel with this work, Zwaan, Briggs, \& Verheijen (2001)
studied a volume limited, optically selected sample of galaxies in the Ursa
Major cluster. The results they obtain are similar those those found here
despite concerns one may have about the optical selection and overdensity in the
region.

Because galaxies with relatively high HI masses contribute most of the HI 
content of the local universe (Rosenberg \& Schneider 2001; Zwaan et al.
1997), and because most luminous spirals have large HI masses, it may seem
reasonable to conclude that luminous spirals must produce most of the
absorption. The failing of this argument is that the correlation
between HI mass and stellar luminosity is quite weak---a large fraction of
high HI-mass systems have very low stellar luminosities, as we show in \S
4.2.

The recent completion of large 21 cm HI surveys provides a galaxy measurement
that is more directly related to HI absorption that can be used to predict 
the characteristics of DLAs. Preliminary studies of these ``blind'' surveys
(Rosenberg \& Schneider 2001; Zwaan et al. 2001) suggested a significantly
larger total HI cross-section for the nearby galaxy population than was
suggested by the earlier optically-selected samples. The present paper
expands on those preliminary results with a more detailed analysis of the
predicted HI and optical properties of the DLAs. In particular, we analyze
near-infrared measurements of our sample of galaxies from the 2-Micron All
Sky Survey (2MASS) to predict the stellar properties of the DLAs.

In Section 2 we show that an HI-selected sample makes a good basis for 
predicting the properties of DLAs. In addition to selecting galaxies
within  the same range of gas column densities, HI mass selection is
nearly equivalent to selection by cross-sectional area. Section 3 contains
a discussion of how $dN/dz$ is calculated for our HI-selected sample and
the resulting value of $dN/dz$ at $z$=0. In Section 4 we discuss our
predictions for the properties of local DLAs: HI masses (\S 4.1), J-band
luminosities (\S 4.2), and kinematics (\S 4.3).

\section{A Surrogate Sample of DLAs in the Local Universe}

Studying the properties of damped \lya\ absorbers is difficult because it
requires the systems to be nearby where absorption-line studies requires
space based UV spectroscopy and the volume surveyed by each line of sight
is small. To study the properties of low-$z$ absorbers we must construct a
surrogate  sample for which we have better statistics. We use a blind, 21
cm emission line  survey, the Arecibo Dual-Beam Survey (ADBS) as a local
comparison sample for  DLA properties. 

The ADBS selected galaxies on the basis of their 21-cm HI flux. The survey
covered $\sim$ 430 deg$^2$ in the Arecibo main beam with a velocity 
coverage of --654 to 7977 \kms\ and  identified 265 galaxies, with HI
masses  ranging from $<2 \times 10^7$ to $>3 \times 10^{10}$ \msolar. Only
a third of the ADBS galaxies had previously identified optical
counterparts, partly because we detected some galaxies in the zone of
avoidance, but also because there is a  substantial population of low
surface brightness galaxies that are not easily identified in
magnitude-limited optical surveys.

To relate the ADBS detection statistics to those of DLA studies, we need
to determine the cross-sectional area for the galaxies above a column
density of $2\times 10^{20}$ cm$^{-2}$. Only a few of the ADBS sources
were resolved in the Arecibo observations and, therefore, only
had a measure of the total HI flux. We made spatially-resolved D-array
follow-up observations from the Very Large Array\footnote{The Very Large
Array is part of the National Radio Astronomy Observatory which is a 
facility of the National Science Foundation operated under cooperative
agreement  by Associated Universities, Inc} (VLA) for 84 of the ADBS
galaxies. For this study, we use the 50 sources with a major axis diameter 
(at 2$\times 10^{20}$ cm$^{-2}$) at least 40\% larger than the beam to 
determine the HI cross-section properties of our sample. 

We find that the HI cross-section is closely correlated with the HI mass
of each source, as shown in Figure \ref{fig:massarea}. The HI mass is
determined  from the source flux, distance, and linewidth using the
standard equation:
\begin{equation}
M_{HI}/M_{\solar} = 2.36 \times 10^5 D^2 \int S \,d v
\end{equation}
where $\int S\,dv$ is the integrated line flux in Jy \kms, and D  is the
distance in Mpc, corrected for large scale velocity flows (Tonry et  al.
2000), adopting H$_0$ = 75 \kms\ Mpc$^{-1}$. The area of the galaxy, 
$A_{DLA} = \pi\,a\cdot b/4$, where $a$ and $b$ are the major and minor axis 
diameters at an HI column density of $2\times 10^{20}$ cm$^{-2}$.
The measurements for each galaxy are provided in the Appendix.

\centerline{\epsfxsize=0.9\hsize{\epsfbox{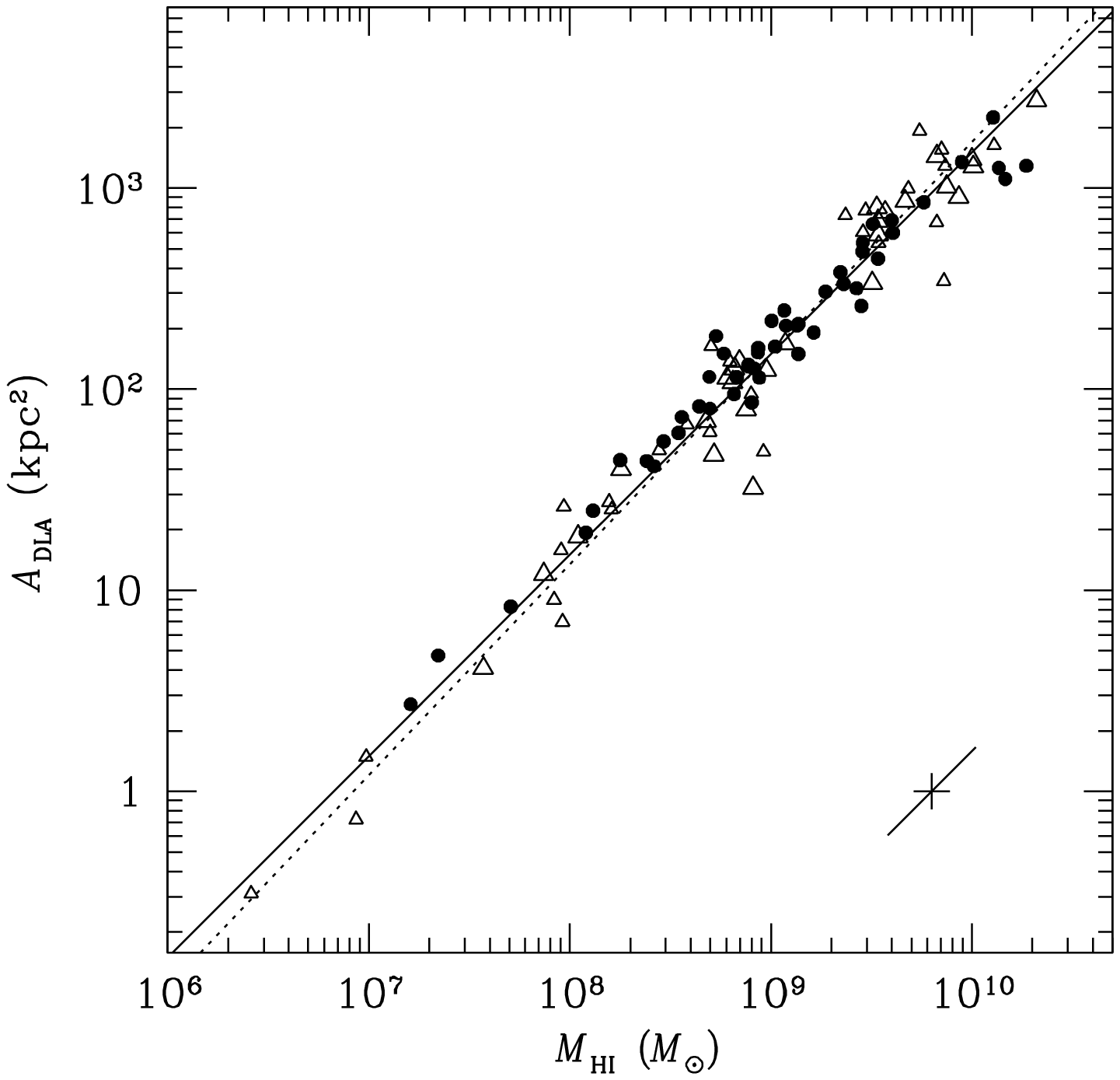}}}
\vspace{0.1in}
\figcaption{\label{fig:massarea} The relationship between the HI cross-section of galaxies above 2
$\times 10^{20}$ cm$^{-2}$ and HI mass. These data are from VLA D-array
measurements of ADBS galaxies (solid circles) and literature values (open
triangles). The galaxies marked with small triangles have minor axes at
least 5 times larger than the beam while the galaxies marked with large
triangles are at least 10 times larger than the beam. The ADBS sources that are
included have a major axis diameter (at 2$\times 10^{20}$ cm$^{-2}$) at least
40\% larger than the beam. The error indicator at the bottom right shows the
effect of $\pm 25$\% distance errors (diagonal line), $\pm 20$\% flux errors,
and $\pm 10$\% radius errors, all generous estimates of the errors.\\ }


Because most of our VLA measurements were not highly resolved, one might
question whether our cross-section estimates are accurate. We have
therefore consulted the literature, collecting high-resolution HI
measurements for a  wide variety of galaxies (see Appendix). The 53
literature sources with minor-axis dimensions at least 5 times larger than
synthesized beam are shown in  Figure \ref{fig:massarea} with open
triangles, and are in excellent agreement with the ADBS sample, shown as
solid circles. The highest-resolution literature sources (with a minor
axis diameter at least 10 times larger than the beam size) are
indicated by large triangles. We find no significant differences between
any of these samples, leading us to conclude that the correlation seen in
Figure \ref{fig:massarea} is unrelated to resolution effects. In any case,
the HI masses are based on the total HI emission from the galaxy, so
resolution effects could only shift points up or down in the figure and
could not artificially generate the observed correlation. Distance errors
{\it do} lead to errors parallel to the observed correlation; however, the
largest conceivable errors lead to small shifts on the scale of this
figure, which spans four orders of magnitude in HI mass.

We averaged forward and backward least squares fits to the high-resolution
data, yielding the relationship:
\begin{equation}
\log(A_{DLA}) = 1.05 \cdot \log(M_{HI}) - 7.24\  ,
\end{equation}
where $A_{DLA}$ is the cross-section in kpc$^2$ and $M_{HI}$ is the
HI mass in solar masses (shown in the figure by the dotted line). We find
similar relationships for the ADBS data alone (although with a slope
slightly less than unity), and for subsets of the high-resolution data.
None of these subsets have slopes  significantly different from 1. If we
combine all of the data, the slope is $1.004 \pm 0.021$, with a
correlation coefficient of 97.7\%. If we assume the relationship has slope
1, the resulting fit is:
\begin{equation}
\log(A_{DLA}) = \log(M_{HI}) - 6.82\  ,
\end{equation}
and the intercept value ($-6.82$) is consistent for the various subsets to
within $\pm 0.03$.

A similar correlation between HI size and HI mass was previously noted 
for high-HI-mass galaxies by Giovanelli \& Haynes (1983) using the
Arecibo  telescope. Their sample included  galaxies in the Virgo Cluster,
yet they found no significant differences between the fits for
HI-deficient and HI-rich galaxies. Verheijen \& Sancisi (2001) likewise
observed a correlation for Ursa Major Cluster  galaxies (with $M_{HI}>10^8
M_{\solar}$ measuring them at a slightly lower HI column density. Our
results demonstrate that the correlation spans a very wide range of
masses, and the correlation is consistent whether the sample is
HI-selected (ADBS galaxies) or optically selected (literature data). This
indicates that the relationship between HI mass and cross-section is
fundamental.

{\it The tightness of this relationship indicates that our HI mass
dependent galaxy selection is nearly equivalent to the cross-sectional
area selection of the DLA population.}

The strong correlation we find between HI cross-section and HI mass is
probably  due, in part, to measuring the galaxies at a high column
density. We expect a larger dispersion at lower column densities because
in the outer parts of galaxies tidal debris and gas-stripping processes
have a strong effect  on the measured extent. For both the ADBS and
literature galaxies, the HI above the $2\times10^{20}$ cm$^{-2}$ isophote is
generally confined to a roughly disk-shaped region, but we see emission
with a wide variety of extents and  morphologies at lower column
densities.

Another reason for the good correlation is that the HI cross-section has a
weak dependence on galaxy inclination. There is a partial cancellation
between the smaller area presented by an inclined disk, and the larger
radius at which the column density remains above $2\times10^{20}$ cm$^{-2}$
because of the longer path length through an inclined disk. We find this
to be demonstrated  empirically for those disk galaxies that authors have
constructed inclination-corrected models of the radial dependence of HI
surface density (see appendix). 
	
Haynes \& Giovanelli (1984) found a similar correlation between HI mass
and the {\it optical} sizes of galaxies for a sample spanning a similar
range in HI mass. We also find that the dimensions of ADBS galaxies as
measured on the Palomar Sky Survey closely correlate with their HI mass.
This suggests that the size of the optically visible region  is closely
related to the dense regions of the HI gas being studied here.

Finally, we note that in comparing emission and absorption line studies we
are comparing gas on very different angular scales: a background quasar's
extent of perhaps 0.25\arcsec-1\arcsec\ for absorption versus
$\sim$10--60\arcsec\ for synthesis array studies. The resolution of the
synthesis data may overlook the small scale properties of these systems. 
If the surface filling factor for N$_{HI} > 2 \times  10^{20}$ cm$^{-2}$
is small, we would overestimate the size of the damped  region for each of
these galaxies. On the other hand, there may be small regions of damped
emission diluted with low column density gas beyond our measured HI
isophote which contribute to a larger HI covering area. Detailed studies
of the  HI covering fraction are available for very few galaxies, and from
what has been done to date, there are disparities in the covering
fractions that have been  derived. Braun \& Walterbos (1992), studying the
small scale HI structure in M31 with a  resolution of 33 pc, find that the
HI surface filling fraction approaches 1 where the emission brightness
temperature exceeds $\sim$5 K (equivalent to a column density of 2 $\times
10^{20}$ cm$^{-2}$ for an asymptotic temperature T$_\infty = 125$ K and a
velocity width of 21.5 \kms). These numbers suggest that the correction
for the covering fraction in these galaxies should be relatively small.

\section{The Total cross-section of DLAs in the Local Universe}

We can estimate the probability of detecting a galaxy in an absorption
line study if we know its cross-section and its relative contribution to
the overall population of absorbers. For each ADBS galaxy we use the
correlation found in the previous section to estimate its cross-section
$A_{DLA}(M_{HI})$, and we determine its relative contribution based on the
relative sensitivity to each source detected. We characterize the latter
by the total volume ${\cal V}_{tot}$ within which each source could have
been detected. The sensitivity criteria and completeness of the ADBS has
been carefully evaluated and tested and is a function of the galaxy's flux
and linewidth, the signal-to-noise ratio of the detection, and properties
of the instrumental response (see Rosenberg \& Schneider 2002 for a
complete discussion). To find the number of systems expected per unit
redshift ($dN/dz$), we sum over all sources:
\begin{equation}
dN/dz = \sum_i 1/{\cal V}_{tot,i} \cdot A_i(M_{HI})\ \cdot (c/H_0) 
\end{equation}
where $1/{\cal V}_{tot}$ gives the volume density for each source.

Thus, summing over all of the ADBS galaxies, our predicted total cross
section at zero redshift is $dN/dz = 0.053 \pm 0.013$. We assume an error
on the ADBS result of 25\% as an estimate of potential systematics in the
mass cross-section relationship and the covering factor to actually
produce damped \lya\ absorption. The formal statistical uncertainty
associated with our estimate is only a few percent.

\centerline{\epsfxsize=0.9\hsize{\epsfbox{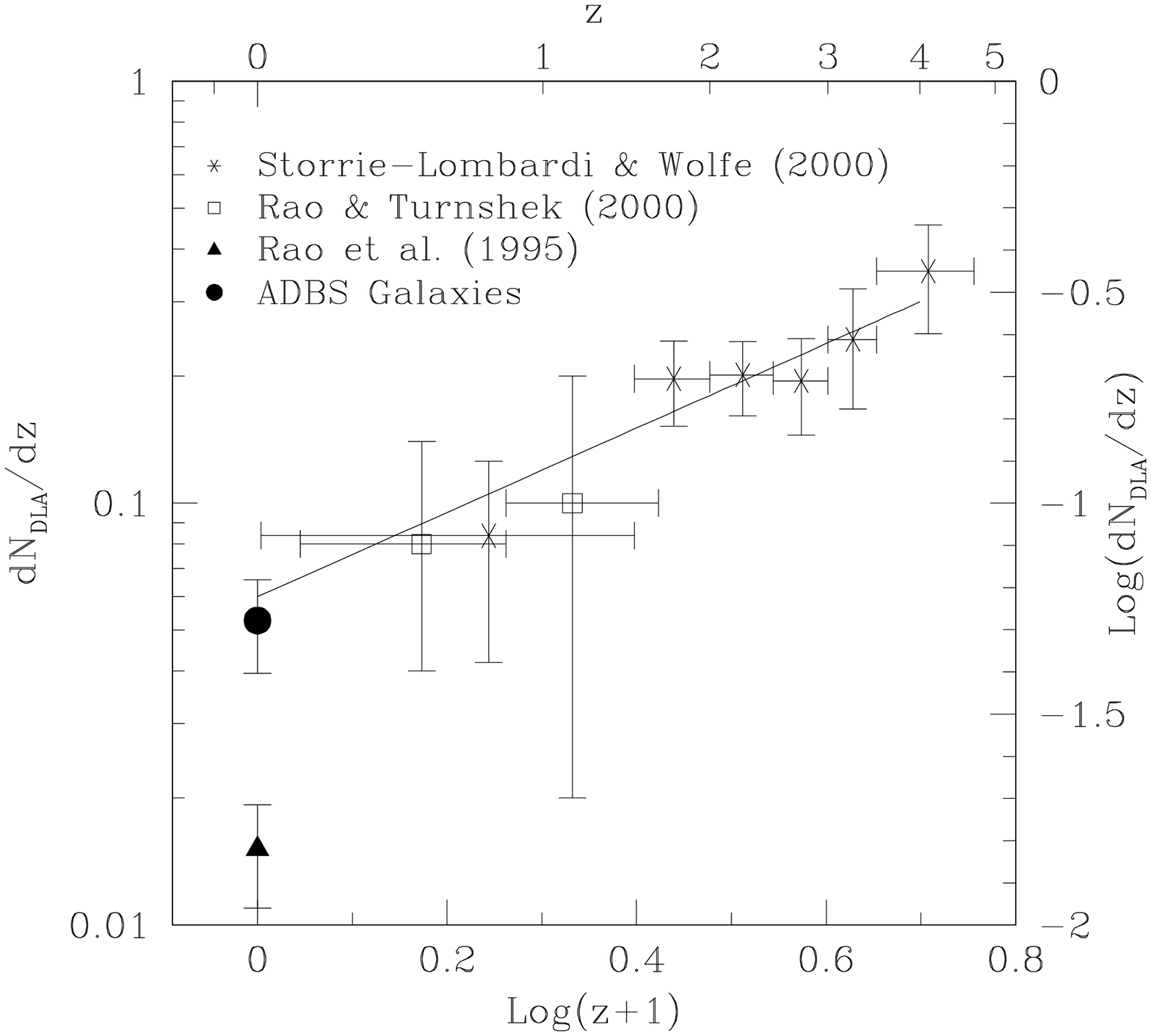}}}
\vspace{0.1in}
\figcaption{\label{fig:dndz} $dN/dz$ of the DLA population as a function of redshift. The
stars are from Storrie-Lombardi \& Wolfe (2000), the open squares are from
Rao \& Turnshek  (2000), the solid triangle is the $z = 0$ point from Rao et
al. (1995), and the filled circle is the $z = 0$ point determined from the
ADBS data in this work. The solid line is a q$_0$ = 0 no evolution model
which indicates that the ADBS sample is consistent with no evolution.\\ }


Figure \ref{fig:dndz} shows $dN/dz$ as a function of redshift for DLAs.
The highest-$z$ points, indicated by stars, are from Storrie-Lombardi
\& Wolfe (2000) who combined their survey of 40 quasars with previous
surveys. The resulting sample contains 646 quasars and 85 DLAs with column
densities $N_{HI}=2\times 10^{20}$ cm$^{-2}$ covering the redshift range
$0.008 < z <  4.694$. The open squares are from Rao \& Turnshek (2000) whose
sample consisted of  12 DLAs detected in 87 Mg II absorber lines of sight.
At $z=0$, the solid triangle is from the study of Rao et al.~(1995), based on
the properties of optically-selected galaxies at low redshift. The filled 
circle is our result described above. Our value for $dN/dz$ at $z=0$ is
consistent with the $dN/dz(z=0.05)$ = $0.08^{+0.09}_{-0.05}$ value
found by Churchill (2001) for Mg II systems with strong Fe II absorption, 
which are good candidates for damped absorption systems (Rao \& Turnshek
2000). These results are also consistent with a similar study of the HI
properties of galaxies by Zwaan et al. (2001). 

Our ADBS result is consistent with the $q_0 = 0$, no evolution model shown
by the solid line in Figure \ref{fig:dndz}. Because of the large error
bars on these  data, we cannot entirely rule out evolution in the DLA
population. However, our result rules out strong evolution of the absorber
population between $z=0$ and $z=0.5$ as was implied by the Rao et al.
(1995) result. These results differ because the $dN/dz$ distribution in
the local universe is not dominated by large disks, but has substantial
contribution from dwarf galaxies as we explore in the next section.

\section{The Predicted Properties of DLAs at $z$ = 0}

We have identified a sample of galaxies that we believe is representative
of DLAs at $z = 0$. We use the properties of this sample to predict the
nature  of the DLA population locally by splitting up the contribution to
$dN/dz$ in Equation 2 according to various properties of the ADBS
galaxies.

\subsection{HI Mass Distribution of DLAs}

Figure \ref{fig:avgarea} shows the contribution of galaxies to $dN/dz$
according to their HI mass. The errors for each bin have been determined
from small number counting statistics using the method described by
Gehrels (1986), and include no estimate of systematic errors. The vertical
line indicates M$_{HI_\ast}$, the knee of the Schechter function fit to the
HI mass function, as determined by Rosenberg \& Schneider (2002). The
dotted histogram corresponds to the determination of HI cross-section from
the averaged forward and backward least squares fit, while the solid line
if from the fit assuming that the slope is unity.

\centerline{\epsfxsize=0.9\hsize{\epsfbox{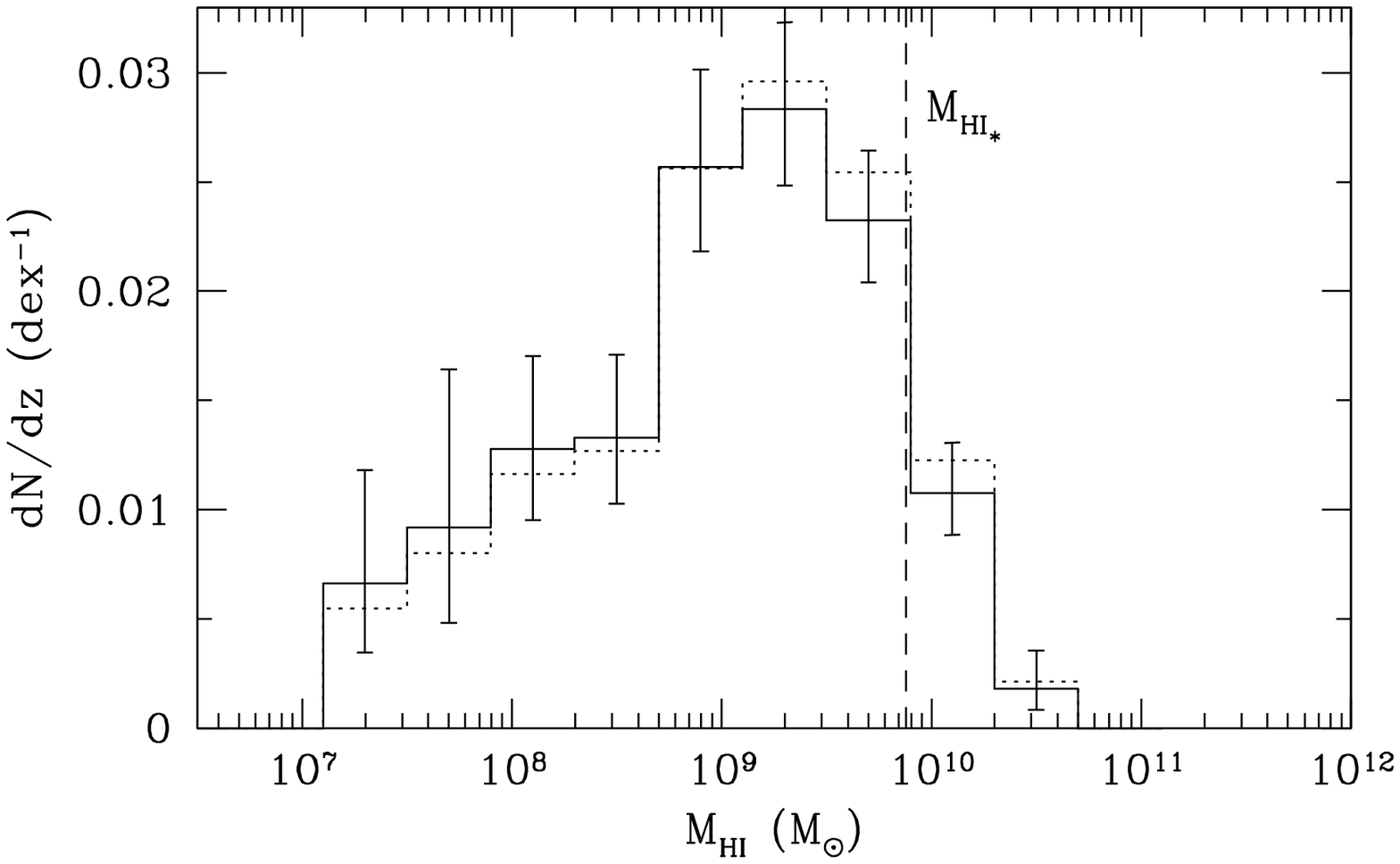}}}
\vspace{0.1in}
\figcaption{\label{fig:avgarea} The value of $dN/dz$ as a function of HI mass for galaxies
identified in the ADBS. The errorbars represent the counting statistics
errors in each bin. The figure demonstrates that most of the contribution
to $dN/dz$ comes from galaxies with HI masses near 10$^9$ \msolar. The
dashed line shows M$_{HI_\ast}$, the HI mass at the knee of the Schechter
fit to the HI mass function (Rosenberg \& Schneider 2002). The Milky Way
contains $\sim 5\times10^9$ \msolar\ of HI (Giovanelli \& Haynes 1988).
The dotted line
corresponds to the determination of HI cross-section from the forward and
backward least squares fit, while the solid line if from the fit assuming
that the slope is one.\\ }


This figure shows that the central 50\% of the cross-sectional area comes
from  galaxies in the range 
$2.9 \times 10^8 < \log M_{HI}/M_\odot < 3.5 \times  10^9$. 
Galaxies above \mstar $ = 7.6 \times 10^9$ \msolar\
account for only 9.5\% of the area. Therefore, this central 50\% of
sources cover a range in HI mass from $0.04 M_{HI_\ast}$ to $0.46 M_{HI_\ast}$, 
relatively low HI mass sources compared to the knee of the
Schechter function.

\subsection{J-Band Magnitude Distribution of DLAs}

Because $L_\ast$ galaxies are a major source of the light in the universe,
and $M_{HI_\ast}$ galaxies are a major source of the HI mass, it is
sometimes assumed that the optical counterparts to the DLAs are HI-rich
$L_\ast$  galaxies -- bright spirals (Wolfe 1995). Instead, many of the
low-$z$ DLAs appear to be low luminosity or are not detected (Colbert \&
Malkan 2001, Steidel et al. 1994, LeBrun et al. 1997, Lanzetta et al.
1997, Rao \& Turnshek 1998, Pettini et al. 2000). Most of the DLA
candidates do not have confirmed redshifts so the actual luminosity
distribution of DLAs is hard to determine. We use the ADBS sample to
predict the distribution of luminosities for DLAs in the local universe by
calculating the contribution to $dN/dz$ in $J$-band luminosity bins.

We obtained the near infrared data used in these analyses from the
2-Micron All-Sky Survey (2MASS, Jarrett et al. 2000). 2MASS used 2
identical 1.3m telescopes in Tucson and Chile which provide simultaneous
$J$, $H$, and $K_s$  observations. Although the observing time per point
is only 7.8 seconds, we were able to detect all but 27 of the galaxies in
our sample on the 2MASS images. Eighteen of these were undetected while
the remaining 9 did not have image data available at the time of this
study. The automated 2MASS galaxy detection algorithms missed 52 faint
galaxies that are visible on the images so we measured them individually
within elliptical isophotes at 21 mag arcsec$^{-2}$. In addition,  13
galaxies had magnitudes measured by the 2MASS galaxy extraction
algorithms, but for various reasons they did not have magnitudes measured
at a $J$-band 21 mag arcsec$^{-2}$ isophote. For the 2 galaxies which have
$J$-band magnitudes measured within the $K_s$-band 20 mag arcsec$^{-2}$
isophote, we substitute these values since these two types of magnitudes
are well correlated. For the remaining 11 galaxies, we use the correlation
between the $J$-band magnitude at 21 mag arcsec$^{-2}$ and the magnitude in
a 10\arcsec\ circular aperture to calculate the magnitude.  (See 
Rosenberg et al. 2002 for details of these galaxy measurements.)

\centerline{\epsfxsize=0.9\hsize{\epsfbox{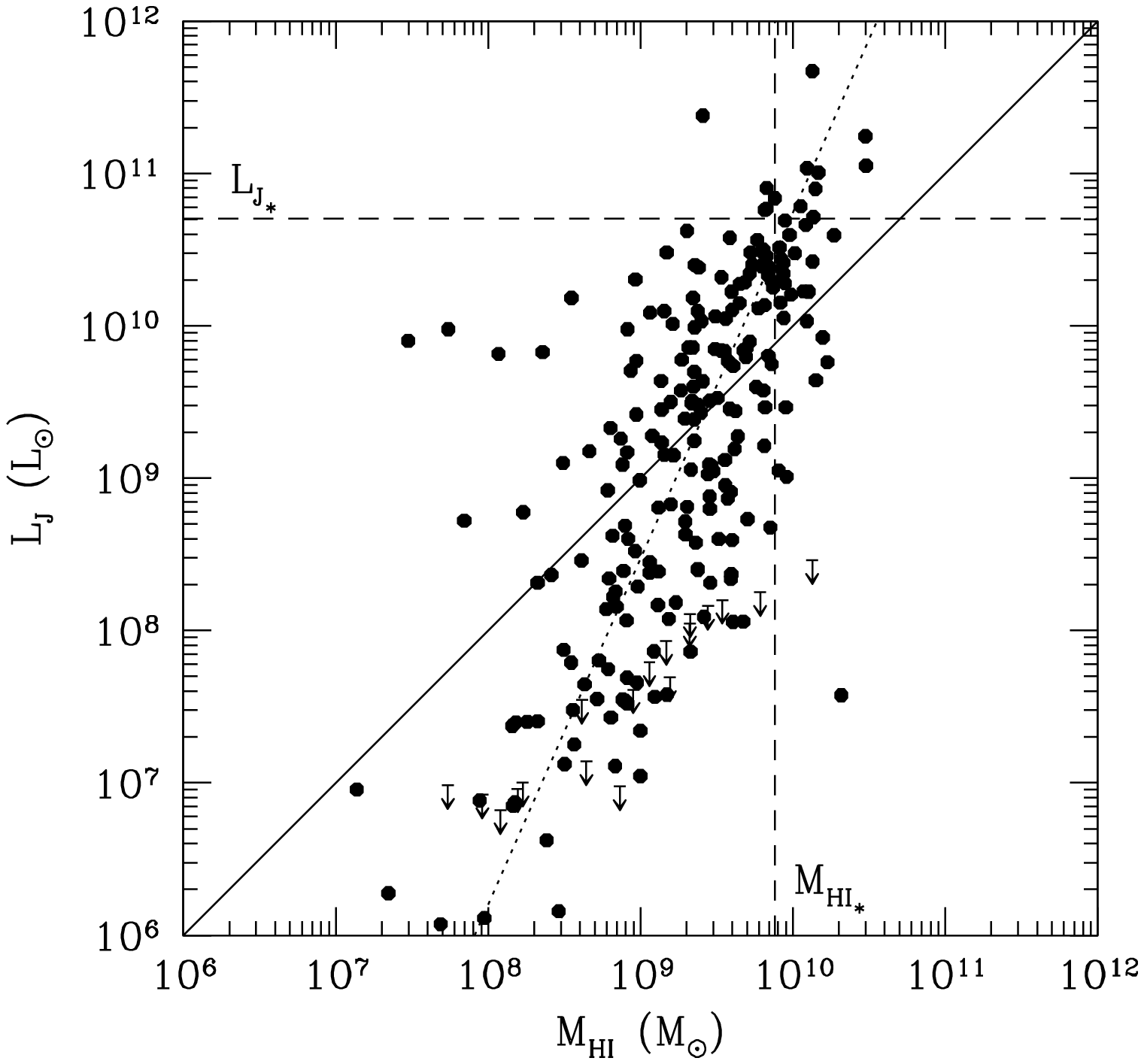}}}
\vspace{0.1in}
\figcaption{\label{fig:massmag} The relationship between $J$-band luminosity and HI mass for
HI-selected galaxies (solid points). The dashed line shows the linear
least squares fit to the points, and, for comparison, the solid line shows
a one-to-one relationship between these quantities. The dotted lines
show $M_{HI_\ast}$ and $L_{J_\ast}$, the knees in the Schechter function
fit.\\ }


The distribution of HI mass as a function of $J$-band luminosity is shown
in Figure \ref{fig:massmag}. The solid line in the figure shows what a
one-to-one relation between $L_J$ and $M_{HI}$ would look like, while the
dashed line shows the actual linear least-squares fit between the values
($\log(L_J)=2.27 \log{M_{HI}}-11.97$).  Although there is a correlation
between the the HI mass and the J-band luminosity, the relationship is
nearly vertical, so that a galaxy with a given HI mass could have a J-band
luminosity spanning four orders of magnitude. There is also a wide scatter
in HI cross-sections for any given J-band luminosity. The dotted lines
show $M_{HI_\ast}$ and $L_{J_\ast}$, the knees in the Schechter function
fit to, respectively, the HI mass function (Rosenberg \& Schneider 2002)
and J-band lumiunosity function (Kochanek et al. 2001).

\centerline{\epsfxsize=0.9\hsize{\epsfbox{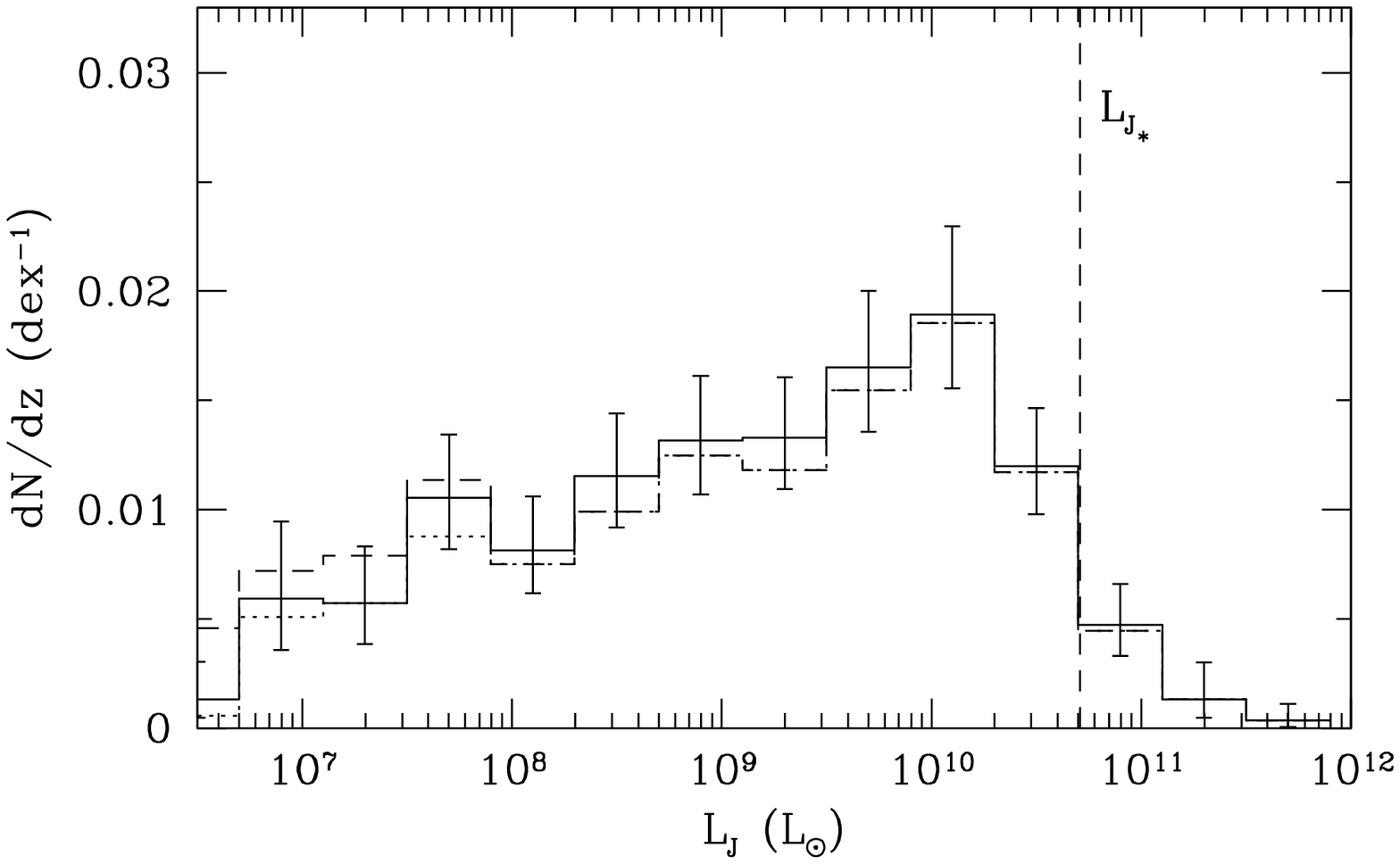}}}
\vspace{0.1in}
\figcaption{\label{fig:dndzj} The value of $dN/dz$ as a function of $L_J$. The function is very
flat, indicating that there is almost uniform probability of intercepting a
galaxy with a luminosity within this range along a given line of sight.
Several alternative histograms are displayed to show the results for
dealing with the 10\% of the ADBS sample for which we had no $J$-band data
(see text). The vertical line shows $L_{J_\ast}$, the knee in the
Schechter function fit to the J-band luminosity function. The total J-band
luminosity of the Milky Way is 5 $\times10^{10} L_\odot$
(Malhotra et al. 1996).\\ }


Figure \ref{fig:dndzj} shows the predicted contribution to $dN/dz$
according to galaxy magnitude along the line of sight. The histograms
assume the unity-slope fit between HI mass and cross-section (see \S 2),
but we show several alternatives for the distribution depending on the
estimates of the $J$-band luminosities of the undetected galaxies (see
below). No matter what assumptions we make about the missing galaxies, the
luminosity distribution is quite flat. Even if we make a conservative
assumption that the missing galaxies are relatively bright, 50\% of
the galaxies lie between luminosities of $8.1\times 10^7 L_\odot$ and
$8.9 \times 10^{9} L_\odot$, spanning a factor of $\sim$110 in
luminosity. Only 4.5\%  of the galaxies are brighter than $J_\ast=
-22.98$ ($L_{J_\ast} = 5.1 \times 10^{10} L_\odot$) (Cole et al. 2001).

In making various alternative assumptions for generating the histograms in
Figure \ref{fig:dndzj}, we have attempted to explore what biases may have
resulted from the 27 galaxies for which we have no 2MASS fluxes. The
dotted-line histogram shows the result when we simply exclude the sources
that were not detected. The solid-line histogram assumes that the missing
galaxies had $J$-band luminosities determined by the fit to the
mass-luminosity relation in Figure \ref{fig:massmag}. This is probably an
upper limit to the luminosities of the galaxies that were too faint to be
detected by 2MASS. The dashed-line histogram assumes that the missing
galaxies have a $J$-band magnitude of 19, slightly fainter than the
faintest apparent magnitude we detect. The differences in the results for
the 3 assumptions is small, However, if the non-detections are all very
faint sources then the distribution is even flatter down to lower J-band
luminosities. 

The definition of ``dwarf" galaxy often varies from paper to paper, but
systems that are fainter than 10\% of $L_\ast$ are often considered
dwarfs. By this definition, 63\% of DLAs in the local universe should be
dwarf galaxies.

\subsection{Kinematics of DLAs}

Prochaska \& Wolfe (1997) argue that the kinematics of DLAs rule out the
possibility that they are dwarf galaxies at the 97\% confidence level
based on simulating dwarfs as slowly rotating hot disk systems. However, their
definition of a dwarf is not a luminosity definition as is often used 
observationally, but instead is a galaxy with circular velocity less than 50
\kms. Our
results above indicate that most DLAs might well be classified as dwarfs in
terms of near infrared luminosity and HI mass.
Significant differences between the kinematics of the low-$z$ ADBS sample,
and the $z > 1.2$ systems studied by Prochaska \& Wolfe (2001, hereafter
PW) could indicate evolution of
the population despite the fairly constant number density of sources found
above.

One complication in comparing the absorption and emission kinematics is
that absorption studies probe a single line-of-sight through a galaxy's
kinematic profile while emission studies probe the full extent of the
system. We can make a comparison between the linewidths for the high and low
redshift systems, but with the difference in the methods of measurement, we will
not be able to show definitively that there has been no evolution of the
population.

PW have analyzed the kinematics of 46 low-ionization lines and 32
high-ionization lines in DLAs and used them to constrain models of the
origin  of these systems (Prochaska \& Wolfe 2001, Prochaska \& Wolfe
1997; Prochaska \&  Wolfe 1998; Wolfe \& Prochaska 2000). PW use the 
low-ionization transitions (Al II, Cr II, Fe II, Ni II, and Si II)  as
tracers of the disk component and the high-ionization transition (C IV) as
a tracer of an infalling component. The velocity width of the system is
defined to be 90\% of the optical depth of the line. We compare our 
emission line  widths to twice
the low-ionization line widths calculated in these studies to account for the 
full rotation of the galaxy. 

\centerline{\epsfxsize=0.9\hsize{\epsfbox{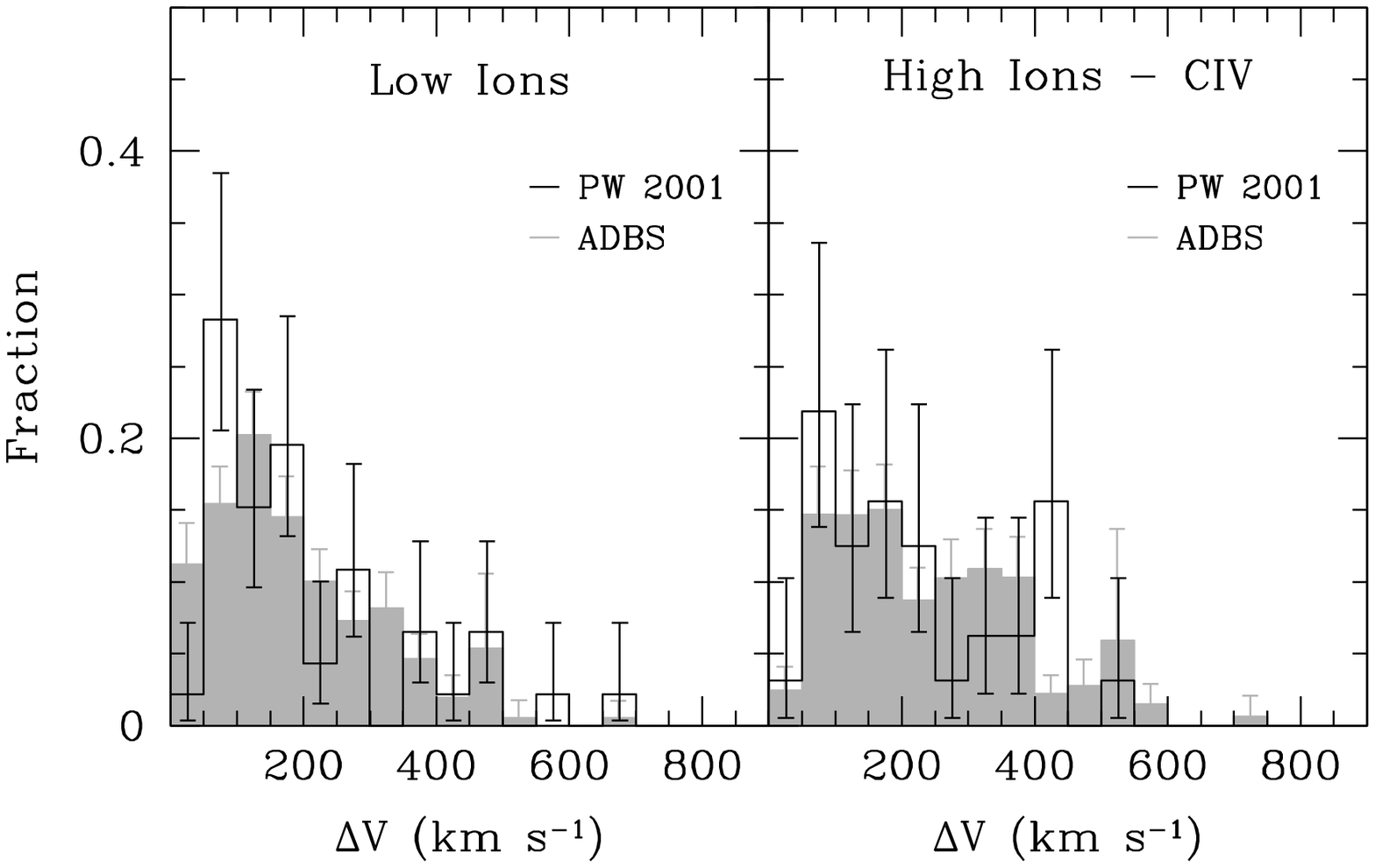}}}
\vspace{0.1in}
\figcaption{\label{fig:dndzw} A comparison of the kinematics of the DLAs the ADBS galaxies. The
figure represents the fraction of the sample in each linewidth bin. The
error bars are derived from the counting statistics for each bin. The top
panels show the comparison between the ADBS data (not inclination corrected;
filled histograms) and twice the low ionization absorption lines widths from
Prochaska \& Wolfe (2001).
The bottom panels shows the comparison between inclination corrected line widths
from the ADBS (filled histograms) and the high ionization lines from this
sample (not multiplied by two).\\ }


\centerline{\epsfxsize=0.9\hsize{\epsfbox{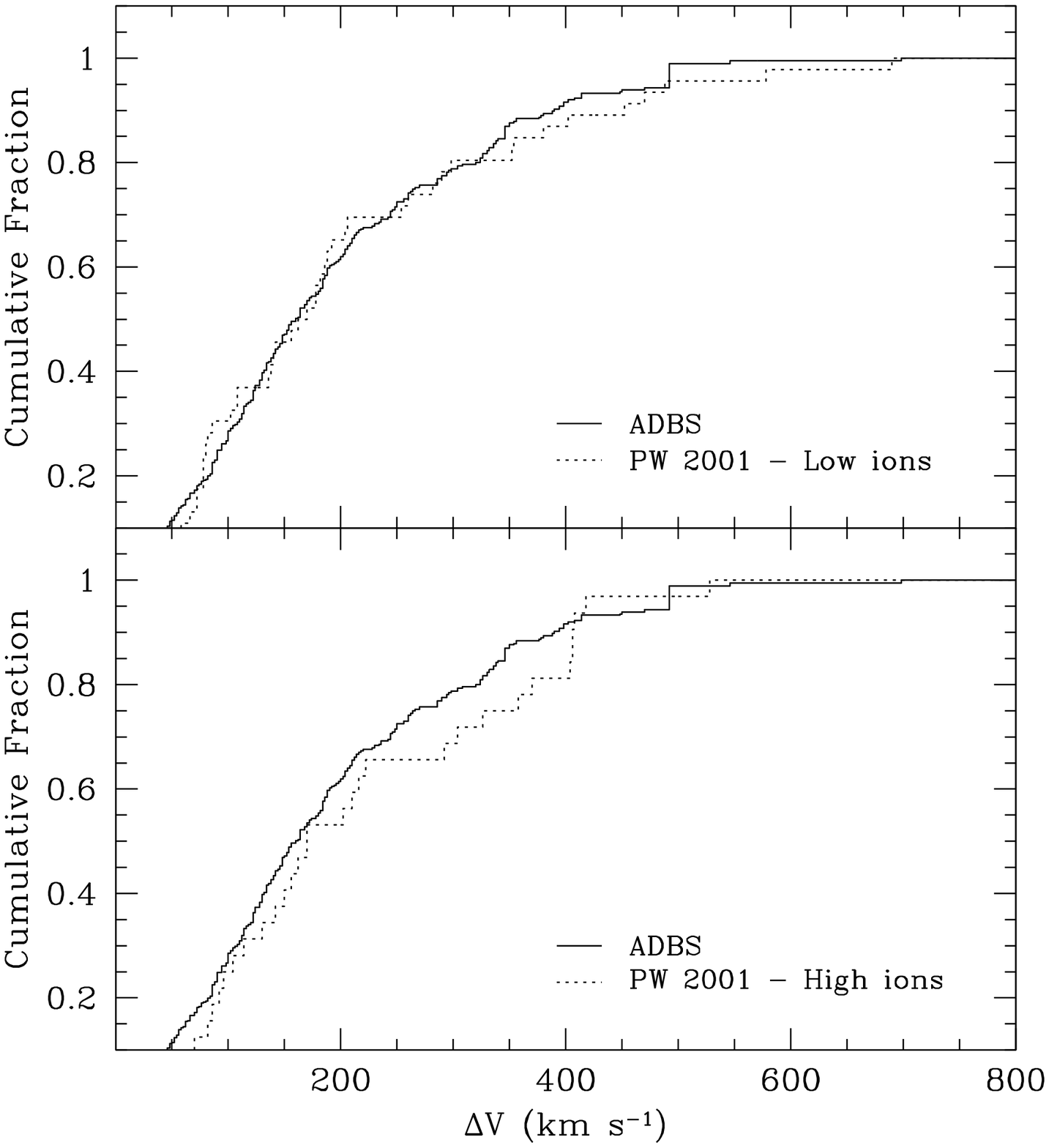}}}
\vspace{0.1in}
\figcaption{\label{fig:dndzwcum} The cumulative fraction of DLAs or galaxies as a
function of linewidth. The top panel shows the comparison between the ADBS data
(not inclination corrected; solid histogram) and twice the low ionization
absorption line widths from Prochaska \& Wolfe (2001; dotted histogram). The
bottom panel shows the comparison
of the inclination corrected ADBS data (solid histogram) with the high
ionization data (not multiplied by two) from Prochaska \& Wolfe (2001; dotted
histogram).\\ }


For the ADBS, we subdivide the calculated $dN/dz$ values (\S 3) according
to the  linewidth (measured at 50\% peak). The left panel in Figure
\ref{fig:dndzw} shows the fraction of sources with each linewidth in the
ADBS sample as compared with the low-ionization species (open histograms) in the 
PW sample (46 DLAs). The
PW widths have been multiplied by two in order to compare the single
sightline velocity dispersion with the full velocity width measured by the HI.
We find that the ADBS velocity widths are not significantly smaller than the DLA
widths as is found for comparison with pure dwarf galaxy samples. We note that
the velocity widths of the DLAs should also be corrected for the effect of only
sampling a small portion of the disk, but we do not know what that correction
is so we do not make it. The upper
panel of Figure \ref{fig:dndzwcum} shows the cumulative distribution comparison
for the same data sets. A KS test indicates that the kinematics can be 
rejected as being from the same sample at the 9\% confidence level. Because we 
do not know how to correct for the differences in disk sampling we can not show 
that there has been no evolution of the population,
but the statistics do not require it to explain the observations. 

The right panel of Figure \ref{fig:dndzw} shows the comparison between the ADBS 
data and the high-ionization species in the PW sample (32 DLAs). Here, we have 
corrected the ADBS
linewidths for inclination and compared then with the high-ionization lines
(which have not been multiplied by two). If the high-ionization species are
infalling, they should be a good probe of the galactic potential, and should
measure gas from both the near and far sides of the galaxy. Here the difference
between the ADBS data and the DLA data would be due to the probes not
penetrating a radial line of sight through the galaxy's center. Figure 
\ref{fig:dndzwcum} shows the cumulative distribution and the KS-test probability
that they are drawn from different samples is 31\% . There is a high line width 
tail for the ADBS
sources that could indicate that there has been evolution, or that the
absorption lines are just not probing the center of the potential.

The K-S test does not preclude that the kinematics of the ADBS galaxies and the 
DLAs are drawn from similar samples. It is not clear whether the kinematic 
differences are due to small number statistics, differences in the way the 
quantities are measured, or evolution of the populations. However,
the ADBS line widths are, in general, as large or larger than the DLA linewidths
indicating that the DLA sample does not require significantly more massive disks
than those identified in an HI selected sample locally.

\section{Summary}

We have predicted the properties of the low-$z$ DLA population from an 
HI-selected galaxy survey of the local universe. The strong linear 
correlation between HI cross-section and HI mass means that selection
effects should be very similar and allows us to use the statistics
available for the properties of ADBS galaxies to predict those of low-$z$
DLAs.

We find a much higher value of $dN/dz$ at $z$=0 from the ADBS than was
found for the optically-selected, spiral dominated sample of Rao et al. 
(1995). The Rao et al.  data have been used as evidence that there was 
evolution
of the DLA population between $z$=0.5 and $z$=0. However, the ADBS
prediction for the total cross-section at $z = 0$ is  consistent with
$dN/dz$ for Mg II absorbers with Fe II absorption, systems that are good
candidates for damped absorption and both of these data sets indicate that
no evolution is required to explain the DLA population.  With the large
error bars on all of the $dN/dz$ points, we can not rule out some
evolution on the DLA population, but it is not required. 

The $dN/dz$ distribution of galaxies as a function of HI mass is peaked
around $10^9$ \msolar, and 50\% of the galaxies should fall within a
factor of $\sim$3 of this value. To date only 2 DLA absorbers have been
studied in emission at 21 cm: one has a M$_{HI}$ of 1.3 $\times 10^9$
\msolar\ (Bowen et al. 2001b), the other was not detected with a 3$\sigma$
upper limit of 2.25 $\times 10^9$ \msolar\ (Kanekar et al. 2001). These
results do not provide a statistical test of our HI  mass distribution,
but the results are consistent with what we would expect. 

The ADBS sample indicates that the stellar luminosities of DLAs should be
distributed fairly evenly over several orders of magnitude. As with the HI
masses, there  are few galaxy/DLA associations for which the galaxy
redshift is confirmed and a luminosity is measured. The absorber
candidates that have been identified  (Steidel et
al. 1994, LeBrun et al. 1997, Lanzetta et al. 1997, Rao \& Turnshek 1998,
Pettini et al. 2000) span a wide range of  luminosities. This range of DLA
luminosities is a warning for high redshift DLA candidate searches---the
absorber candidates are not necessarily the nearest L$_\ast$ galaxy to the
line of sight. 

We have compared the internal kinematics of the ADBS galaxies with those
for high redshift DLAs (a low redshift sample does not exist). Prochaska
\& Wolfe (1997), through simulations of gas disks, rule out galaxies with small
velocity widths
as the primary  source of damped emission at a 97\% confidence level. The
ADBS galaxies that we  associate with the DLAs locally have larger line widths
than the Prochaska \& Wolfe ``dwarf" galaxies, but the sample consists of a range
of galaxy types dominated by galaxies with HI masses of $\sim$0.1 \mstar\ and 
luminosities below L$_{J_\ast}$. We find that we can not rule
out evolution in the kinematics of HI-rich galaxies, but it is not required 
and more massive
HI disks then found in our local sample are not necessary to account for the
kinematics. 

\acknowledgements
We thank Jason X. Prochaska for helpful discussions about the kinematics 
of DLAs and the referee for a thoughtful reading of this paper. We would also
like to thank the VLA staff for their assistance in the observations and
reduction of this data, and the 2MASS team for all of the hard work that went
into making the survey a reality.

\appendix
\section{Appendix}

In this appendix we present VLA data for the ADBS sample of galaxies,
as well as high-resolution literature data used as a comparison sample
to confirm our results.  In addition, we provide some further analysis
of the HI cross-sections of the galaxies beyond the analysis in the
main text.

\subsection{VLA HI Data for the ADBS}

The galaxies in the ADBS were observed with the VLA in D-array, with
relatively short duration (10 to 15 min) integrations. These were
usually split into two time intervals in order to improve the $u$-$v$
coverage of the observations. Further details of the observational
procedures can be found in Rosenberg \& Schneider (2002).

We present the basic information on the 50 sources detected with the
VLA that have a major axis diameter (at $2 \times 10^{20}$ cm$^{-2}$) at 
least 40\% larger than the beam.. All were reduced using the 
observatory's {\it AIPS} routine
{\it IMAGR}, setting the {\it ROBUST} parameter to 0. This provides a
good compromise, providing nearly as good noise characteristics
as natural weighting and nearly as small a beam as uniform weighting.
Table A.1 lists the FWHM dimensions of the beam along with other measured
parameters for the galaxies.

The elliptical dimensions of the galaxies at a column density of
$2\times10^{20}$ cm$^{-2}$ were fit to the total HI maps using the {\it
IRAF STSDAS} package {\it ELLIPSE}. The mean velocities and total HI
masses determined from the VLA observations are all consistent with our
Arecibo detection measurements. We also provide a flow-corrected distance
estimate based on Tonry et al. (2000), assuming $H_0=75$ km s$^{-1}$ 
Mpc$^{-1}$.

\subsection{HI Synthesis Data from the Literature}

We compiled higher-resolution HI measurements from the literature for
comparison with our results. We consulted sources listed in Martin's
(1998) bibliographic compilation of HI maps in order to identify synthesis
observations with beam sizes at least ten times smaller than the reported
HI dimensions. We further restricted our sample to sources that Martin
indicated reached column densities below $2\times10^{20}$ cm$^{-2}$ and
which had both HI contour maps and radial density profiles. This yielded
about 100 candidate sources. We found that the resulting list contained
few low-mass sources, so we added several more-recent papers that
concentrated on dwarf galaxies. The resulting sample spans all galaxy
types from giant ellipticals to dwarf irregulars.

Table A.2 compiles the information collected from these papers in a similar
format as for our ADBS galaxies. None of the papers directly reported the
dimensions of the galaxy at $2\times10^{20}$ cm$^{-2}$, so we estimated
them from the HI column density contour maps. The beam dimensions are
those reported in each paper for the contour maps. In a few cases (noted
by the word ``ring'' or ``mult'' in the notes column of the table) the
distribution of HI above $2\times10^{20}$ cm$^{-2}$ was not well described
as a disk. In these cases, we give major and minor axis dimensions that
approximately indicate (a) the combined length of the regions of HI that
exceeded our fiducial level, and (b) the narrower dimension of these
regions. These dimensions should yield a fairly accurate total
cross-section for the sources. We note that many other sources also showed
irregularities around the edges, or small holes or small regions outside
the main disk, but these amount to only slight deviations in the
overall dimensions of the sources.

For papers that displayed inclination-corrected, azimuthally-averaged
radial density profiles, we again used the published plots to estimate the
radius at which the HI column density reached our fiducial level. In a few
cases the radial profile was plotted, but did not exceed the
$2\times10^{20}$ cm$^{-2}$ level.  In all of these cases, the HI contour
map had regions above this level, but the azimuthal averaging had produced
a lower mean level. We note that these radial profiles are quite sensitive
to the inclination estimates. As a result, they tend to be more accurate
for disk galaxies, and much less so for dwarf irregulars.

The redshifts and morphological types listed in the table were taken from
the NASA/IPAC Extragalactic Database (NED). The distances are again based
on the flow-corrected Tonry model, except when the redshift was less than
250 km s$^{-1}$, in which case we used the distance quoted by the author.
For two of these nearby galaxies, UGC 8091 (Dohm-Palmer et al. 1998) and
UGC 12613 (Gallagher et al. 1998), recent distance estimates were
significantly different from those reported in the HI paper, and we have
adopted the newer values.

\subsection{The HI Mass/Cross-Section Correlation}

The strong, one-to-one correlation between HI mass and cross-section over
such a wide range of galaxy sizes and masses is somewhat surprising. At
first glance, it suggests that all types of galaxies share the same
surface density of HI. Actually the data are a little more complicated to
interpret, since the relationship is between the total HI, and the area of
the disk that exceeds the relatively high surface density of
$2\times10^{20}$ cm$^{-2}$. If all galaxies had their HI in exponential
disks with varying scale lengths but the same central surface brightness,
such a relationship would pertain, but clearly that is not a realistic
description of galaxy HI distributions. We present here a few exploratory
comments about the data presented in Table A.2 to examine this correlation
in greater detail.

\begin{figure}[ht]
\plotone{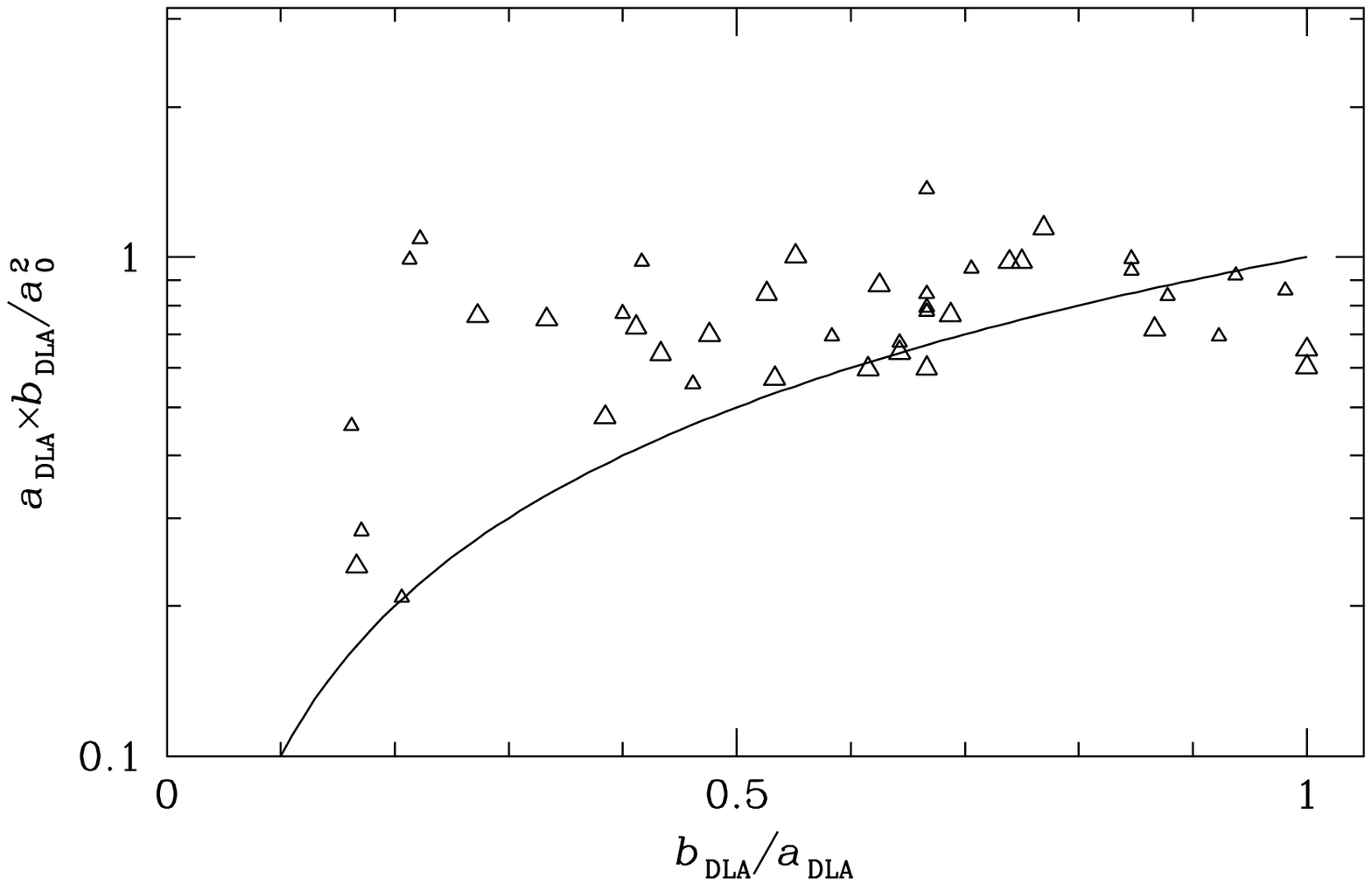}
\caption{The ratio of the observed cross-section to the predicted face-on
cross-section as a function of observed axis ratio. The galaxies marked
with small triangles have minor axes at least 5 times larger then the beam
while the galaxies marked with large triangles are at least 10 times
larger than the beam. The curve shows the expected ratio if the 
cross-section varied as cos$(i)$.}
\label{fig:axrat}
\end{figure}

For the central purpose of this paper, we focused on the relationship
between the HI mass and the {\it observed} cross-section of HI on the sky.
However, as we briefly noted in \S 2, the observed cross-section is
relatively independent of inclination. We show this in
Figure \ref{fig:axrat}, where we plot the ratio of observed to face-on cross
sections above $2\times10^{20}$ cm$^{-2}$ as a function of the observed
axis ratio. The sample shown is the subset of galaxies from
Figure \ref{fig:massarea} that also had inclination-corrected radial profile
models. The cross-section does not simply decline as $\cos(i)$ (curve in
figure), as would occur for a sharp-edged disk. Because of the longer path
length through the disk, the $2\times10^{20}$ cm$^{-2}$ isophote shifts
outward to a physically larger radius by an amount that depends on the
scale-length of the HI distribution. The two effects nearly cancel for
most inclinations, although edge-on galaxies do finally exhibit a
significantly smaller cross-section. Overall, the observed cross-sections
are about 25\% smaller, on average, than the inclination-corrected models
predict for galaxies with axis ratios larger than $b_{DLA}/a_{DLA}>0.5$.

The scatter about the fit between mass and cross-sectional area improves
slightly when we use azimuthally-averaged, inclination-corrected model
results. Using these face-on cross-sections yields a similar slope
($1.04\pm0.04$) as we found using the observed cross-sections in a linear
least squares fit, although the intercept shifts because of the larger
cross-section found on average for the face-on models. We will assume both
actually have slope unity so that we can just compare the distribution of
$\log[M_{HI}/A_{DLA}]$. For the 64 galaxies with measurements of both the
face-on, $a_0$, and observed dimensions, $a_{DLA} \times b_{DLA}$, the rms 
scatter of $\log[M_{HI}/A_{DLA}]$ is 0.194 and 0.227 respectively.

There seems to be little dependence with morphological type. If we use
observed cross-sections for galaxies which are not highly inclined
($b_{DLA}/a_{DLA}>0.5$), we find the following results by morphological type:

\begin{table}[h]
\begin{tabular}{ccc}
Type & N & $\log[M_{HI}/A_{DLA}]$ \\
\hline
Sa -- Sb & 10    & $6.71\pm0.08$ \\
Sbc -- Sd & 14    & $6.85\pm0.03$ \\
Sdm -- Im & 15    & $6.77\pm0.04$ \\
\end{tabular}
\end{table}

The earliest-type galaxies (Sa -- Sb) may have slightly lower values of
$\log[M_{HI}/A_{DLA}]$ than late-types, but the difference is marginal. This
appears to be true for the S0's as well, but our few examples have
peculiar HI distributions, so any conclusions depend on whether the model
or observed cross-sections are used, and whether outliers are rejected.
While there may be some dependence on morphological type, the remarkable
thing is how little variability there is between types over an enormous
range of masses.

\clearpage
\begin{deluxetable}{lc@{ }c@{ }c@{}c@{ }c@{ }rrr}
\tablewidth{4.9in}
\tablenum{A.1}
\tablecaption{HI sizes at N$_{HI} = 2 \times 10^{20}$ from the ADBS}
\tablehead{
\colhead{Name} & \colhead{D$_{beam}$} & \colhead{D} 
& \colhead{V$_{hel}$} & \colhead{Dist.} & \colhead{$\log$(M$_{HI}$)} \\
\colhead{} & \colhead{[\arcm]} & \colhead{[\arcm]} 
& \colhead{} & \colhead{[Mpc]} & \colhead{[M$_\odot$]} }
\startdata
002526+2136 &  0.95$\,\times\,$0.70 &  \ \    1.5$\,\times\,$1.1   &    5001 &   66.7 &   9.44 \\
014729+2719 &  0.87$\,\times\,$0.81 &  \ \    7.8$\,\times\,$2.9   &     743 &    9.9 &   9.14 \\
015434+2312 &  0.98$\,\times\,$0.87 &  \ \    1.5$\,\times\,$1.4   &    5298 &   70.6 &   9.60 \\
020320+1837 &  0.90$\,\times\,$0.83 &  \ \    2.8$\,\times\,$2.7   &    2727 &   36.4 &   9.50 \\
020320+2345 &  0.88$\,\times\,$0.82 &  \ \    1.5$\,\times\,$1.4   &    3081 &   41.1 &   9.06 \\
020405+2412 &  0.95$\,\times\,$0.85 &  \ \    1.5$\,\times\,$1.5   &     989 &   13.2 &   8.11 \\
022859+2808 &  0.87$\,\times\,$0.82 &  \ \    2.5$\,\times\,$1.5   &    1341 &   17.9 &   8.69 \\
023323+2810 &  0.87$\,\times\,$0.82 &  \ \    1.8$\,\times\,$1.2   &    1316 &   17.5 &   8.23 \\
023622+2526 &  0.88$\,\times\,$0.83 &  \ \    9.6$\,\times\,$2.3   &     987 &   13.2 &   9.45 \\
025537+1938 &  0.95$\,\times\,$0.87 &  \ \    3.1$\,\times\,$1.5   &     893 &   11.9 &   8.38 \\
025726+1008 &  0.94$\,\times\,$0.86 &  \ \    3.0$\,\times\,$1.7   &    1092 &   14.6 &   8.56 \\
030546+2212 &  0.96$\,\times\,$0.87 &  \ \    3.1$\,\times\,$1.9   &    4365 &   58.2 &   9.94 \\
031420+2409 &  0.88$\,\times\,$0.84 &  \ \    3.9$\,\times\,$1.9   &    1523 &   20.3 &   9.12 \\
040344+2209 &  0.99$\,\times\,$0.90 &  \ \    2.2$\,\times\,$2.2   &    6307 &   84.1 &  10.09 \\
040411+2207 &  0.97$\,\times\,$0.87 &  \ \    1.8$\,\times\,$1.5   &    6320 &   84.3 &  10.13 \\
053017+2233 &  0.92$\,\times\,$0.83 &  \ \    1.5$\,\times\,$1.0   &    2555 &   34.1 &   8.64 \\
055517+2526 &  0.90$\,\times\,$0.84 &  \ \    1.8$\,\times\,$1.6   &    1547 &   20.6 &   8.64 \\
062054+2008 &  1.10$\,\times\,$0.89 &  \ \    3.7$\,\times\,$1.8   &    1279 &   17.1 &   8.86 \\
065406+0834 &  1.21$\,\times\,$0.93 &  \ \    2.6$\,\times\,$1.8   &    2420 &   32.3 &   9.42 \\
070920+2038 &  1.09$\,\times\,$0.90 &  \ \    2.6$\,\times\,$1.8   &    4879 &   65.1 &  10.27 \\
071225+2342 &  1.12$\,\times\,$0.89 &  \ \    2.3$\,\times\,$1.9   &    4138 &   55.2 &   9.76 \\
071352+1031 &  1.02$\,\times\,$0.81 &  \ \    3.6$\,\times\,$2.1   &     304 &    4.1 &   7.69 \\
071553+1207 &  1.04$\,\times\,$0.84 &  \ \    2.4$\,\times\,$2.0   &    1972 &   26.3 &   8.98 \\
081726+2110 &  1.22$\,\times\,$0.89 &  \ \    1.8$\,\times\,$1.7   &    2036 &   27.1 &   8.92 \\
090024+2536 &  1.27$\,\times\,$0.87 &  \ \    3.8$\,\times\,$2.3   &    1806 &   24.1 &   9.36 \\
090526+2533 &  1.09$\,\times\,$0.88 &  \ \    2.6$\,\times\,$2.1   &    2869 &   38.3 &   9.44 \\
101421+2207 &  1.05$\,\times\,$0.86 &  \ \    1.5$\,\times\,$1.2   &    1686 &   22.5 &   8.49 \\
103143+2518 &  1.22$\,\times\,$0.86 &  \ \    3.7$\,\times\,$2.6   &    1205 &   16.1 &   9.00 \\
103439+2305 &  1.20$\,\times\,$0.85 &  \ \    2.1$\,\times\,$1.7   &    1142 &   15.2 &   8.46 \\
104722+1404 &  1.10$\,\times\,$0.84 &  \ \    2.5$\,\times\,$2.3   &     264 &    3.5 &   7.34 \\
104916+1226 &  1.39$\,\times\,$0.90 &  \ \    4.1$\,\times\,$2.5   &    1238 &   16.5 &   8.73 \\
110710+1834 &  1.29$\,\times\,$0.87 &  \ \    4.0$\,\times\,$2.1   &    1082 &   14.4 &   8.94 \\
112914+2035 &  1.01$\,\times\,$0.85 &  \ \    2.1$\,\times\,$2.0   &    1392 &   18.6 &   8.82 \\
115004+2628 &  1.09$\,\times\,$0.86 &  \ \    2.0$\,\times\,$1.8   &    2096 &   27.9 &   9.21 \\
115040+2531 &  1.07$\,\times\,$0.86 &  \ \    2.3$\,\times\,$1.3   &    2149 &   28.7 &   8.91 \\
120033+2004 &  1.04$\,\times\,$0.85 &  \ \    3.9$\,\times\,$2.9   &    1254 &   16.7 &   9.14 \\
121516+2038 &  1.04$\,\times\,$0.85 &  \ \    5.0$\,\times\,$4.1   &     743 &    9.9 &   8.89 \\
123606+2602 &  1.51$\,\times\,$0.85 &  \ \   15.0$\,\times\,$3.0   &    1453 &   19.4 &  10.17 \\
124516+2708 &  1.66$\,\times\,$0.86 &  \ \    3.0$\,\times\,$2.2   &    1050 &   14.0 &   8.87 \\
125223+2138 &  1.29$\,\times\,$0.87 &  \ \    3.1$\,\times\,$2.3   &     479 &    6.4 &   8.08 \\
131652+1232 &  1.09$\,\times\,$0.85 &  \ \    3.3$\,\times\,$2.2   &     689 &    9.2 &   8.41 \\
141556+2303 &  0.93$\,\times\,$0.85 &  \ \    2.7$\,\times\,$2.2   &     195 &    2.6 &   7.21 \\
144842+1226 &  1.18$\,\times\,$0.84 &  \ \    3.3$\,\times\,$2.6   &    1950 &   26.0 &   9.35 \\
145647+0930 &  1.35$\,\times\,$0.84 &  \ \    2.4$\,\times\,$2.0   &    3221 &   42.9 &   9.59 \\
154540+2805 &  1.09$\,\times\,$0.86 &  \ \    2.4$\,\times\,$1.9   &    2353 &   31.4 &   9.27 \\
163122+2010 &  1.28$\,\times\,$0.85 &  \ \    3.5$\,\times\,$1.6   &    2560 &   34.1 &   9.53 \\
214731+2209 &  0.99$\,\times\,$0.84 &  \ \    1.9$\,\times\,$1.4   &    2033 &   27.1 &   8.90 \\
214813+2209 &  1.00$\,\times\,$0.84 &  \ \    2.3$\,\times\,$2.0   &    1967 &   26.2 &   9.06 \\
225557+2610 &  0.96$\,\times\,$0.84 &  \ \    1.4$\,\times\,$1.1   &    2979 &   39.7 &   8.72 \\
230433+2709 &  0.96$\,\times\,$0.85 &  \ \    3.0$\,\times\,$1.6   &    1439 &   19.2 &   8.82 \\
\enddata
\end{deluxetable}

\clearpage
\begin{deluxetable}{lrrc@{ }rc@{ }c@{ }lll}
\rotate
\small
\tablewidth{9.1in}
\tablenum{A.2}
\tablecaption{HI sizes at N$_{HI} = 2 \times 10^{20}$ from the Literature}
\tablehead{
\colhead{Name} & \colhead{D$_{beam}$} & \colhead{D} & \colhead{D$_{corr}$} & \colhead{V$_{hel}$}
& \colhead{Dist.} & \colhead{$\log$(M$_{HI}$)} & \colhead{Type} & \colhead{Morph.}
& \colhead{Reference}\\
\colhead{} & \colhead{[\arcm]} & \colhead{[\arcm]} & \colhead{[\arcm]}
& \colhead{} & \colhead{[Mpc]} & \colhead{[M$_\odot$]} & \colhead{}
& \colhead{} & \colhead{}}
\startdata
NGC 55 &       0.75$\,\times\,$0.75 &   41.0$\,\times\,$7.0   &  31.9 &    129 &   1.6 &  8.96 &  SB(s)m:         &\nodata  &  Puche~et~al.~(1991) \\     
NGC 247 &      0.75$\,\times\,$0.75 &   37.0$\,\times\,$6.0   &  22.0 &    160 &   2.5 &  8.90 &  SAB(s)d         &\nodata  &  Carignan~\&~Puche~(1990b) \\
NGC 300 &      0.83$\,\times\,$0.83 &   32.0$\,\times\,$20.0  &  27.0 &    144 &   1.8 &  8.84 &  SA(s)d          &\nodata  &  Puche~et~al.~(1990) \\    
IC 1613 &      2.00$\,\times\,$2.00 &   22.0$\,\times\,$4.0   & \nodata &   -234 &   0.7 &  7.78 &  IB(s)m          & ring  &  Lake~\&~Skillman~(1989) \\ 
UGC 891 &      0.64$\,\times\,$0.56 &    4.5$\,\times\,$1.7   &   5.1 &    643 &  14.6 &  9.11 &  SABm:           &\nodata  &  van~Zee~et~al.~(1997) \\   
M 74 &         0.80$\,\times\,$0.39 &   15.0$\,\times\,$13.0  &  16.5 &    657 &  14.5 & 10.32 &  SA(s)c          &\nodata  &  Wevers~et~al.~(1986) \\    
UGCA 20 &      0.18$\,\times\,$0.18 &    4.5$\,\times\,$1.5   & \nodata &    498 &  12.2 &  8.58 &  Im              & mult  &  van~Zee~et~al.~(1996) \\   
NGC 925 &      0.77$\,\times\,$0.42 &   12.0$\,\times\,$8.0   &  12.7 &    553 &  11.9 &  9.93 &  SAB(s)d         &\nodata  &  Wevers~et~al.~(1986) \\    
NGC 1058 &     0.72$\,\times\,$0.70 &   10.5$\,\times\,$7.0   &   7.3 &    518 &  11.1 &  9.46 &  SA(rs)c         &\nodata  &  Dickey~et~al.~(1990) \\    
NGC 1073 &     0.34$\,\times\,$0.33 &    7.0$\,\times\,$7.0   &   8.7 &   1211 &  21.0 &  9.82 &  SB(rs)c         &\nodata  &  England~et~al.~(1990) \\   
UGC 2259 &     0.37$\,\times\,$0.37 &    4.1$\,\times\,$3.6   &   4.2 &    583 &  11.8 &  8.80 &  SB(s)dm         &\nodata  &  Carignan~et~al.~(1988) \\  
NGC 1140 &     0.62$\,\times\,$0.50 &    7.7$\,\times\,$3.0   &   4.7 &   1501 &  25.1 &  9.72 &  IBm:            &\nodata  &  Hunter~et~al.~(1994) \\    
NGC 1291 &     0.80$\,\times\,$0.80 &    8.5$\,\times\,$2.5   & \nodata &    839 &  15.5 &  9.41 &  (R)SB(l)0/a     & ring  &  van~Driel~et~al.~(1988b) \\
NGC 1300 &     0.33$\,\times\,$0.33 &    7.0$\,\times\,$4.5   &   7.0 &   1568 &  25.7 & 10.00 &  (R')SB(s)bc     &\nodata  &  England~(1989) \\          
UGC 2684 &     0.50$\,\times\,$0.48 &    3.5$\,\times\,$1.4   &   3.1 &    350 &   8.9 &  8.38 &  Im?             &\nodata  &  van~Zee~et~al.~(1997) \\   
NGC 1343 &     1.17$\,\times\,$0.78 &    6.3$\,\times\,$4.9   & \nodata &   2215 &  31.1 &  9.72 &  SAB(s)b:        &\nodata  &  Taylor~et~al.~(1994) \\    
NGC 1398 &     0.99$\,\times\,$0.96 &    9.0$\,\times\,$6.0   &   8.3 &   1407 &  23.2 &  9.74 &  (R')SB(rs)ab    &\nodata  &  Moore~\&~Gottesman~(1995) \\
IC 342 &       2.00$\,\times\,$1.90 &   60.0$\,\times\,$50.0  &  52.0 &     31 &   2.0 &  9.53 &  SAB(rs)cd       &\nodata  &  Newton~(1980) \\           
NGC 1560 &     0.23$\,\times\,$0.22 &   18.0$\,\times\,$3.0   &  15.0 &    -36 &   3.0 &  8.91 &  SA(s)d          &\nodata  &  Broeils~(1992) \\          
UGC 3174 &     0.72$\,\times\,$0.62 &    4.8$\,\times\,$2.4   &   4.1 &    670 &  10.9 &  8.60 &  IAB(s)m:        &\nodata  &  van~Zee~et~al.~(1997) \\   
UGCA 116 &     0.14$\,\times\,$0.13 &    6.1$\,\times\,$1.3   &   2.8 &    789 &  10.8 &  8.70 &  BCD/Sbc         & mult  &  van~Zee~et~al.~(1998) \\   
NGC 2366 &     0.43$\,\times\,$0.42 &   13.0$\,\times\,$5.0   &  11.7 &    100 &   3.3 &  8.72 &  IB(s)m          &\nodata  &  Wevers~et~al.~(1986) \\    
NGC 2403 &     0.75$\,\times\,$0.75 &   30.0$\,\times\,$16.0  &  29.0 &    131 &   3.2 &  9.51 &  SAB(s)cd        &\nodata  &  Wevers~et~al.~(1986) \\    
UGC 3974 &     0.13$\,\times\,$0.12 &    8.0$\,\times\,$6.0   &   7.0 &    272 &   3.5 &  8.25 &  IB(s)m          &\nodata  &  Walter~\&~Brinks~(2001) \\ 
UGC 4483 &     0.18$\,\times\,$0.15 &    2.6$\,\times\,$2.0   &   2.1 &    178 &   3.4 &  7.57 &  Dwarf           &\nodata  &  van~Zee~et~al.~(1998) \\   
NGC 2787 &     0.45$\,\times\,$0.42 &    6.0$\,\times\,$1.0   & \nodata &    696 &  10.9 &  8.73 &  SB(r)0+         & ring  &  Shostak~(1987) \\          
NGC 2841 &     1.08$\,\times\,$0.85 &   27.0$\,\times\,$6.0   &  12.2 &    638 &   8.3 &  9.37 &  SA(r)b:         &\nodata  &  Bosma~(1981) \\            
NGC 2903 &     0.55$\,\times\,$0.55 &   17.0$\,\times\,$7.0   &  12.8 &    556 &   4.6 &  9.07 &  SB(s)d          &\nodata  &  Wevers~et~al.~(1986) \\    
Holmberg I &   0.20$\,\times\,$0.18 &    5.0$\,\times\,$4.3   & \nodata &    143 &   3.6 &  8.04 &  IAB(s)m         &\nodata  &  Ott~et~al.~(2001) \\       
Leo A &        0.38$\,\times\,$0.38 &    8.7$\,\times\,$4.3   & \nodata &     20 &   2.2 &  7.87 &  IBm             &\nodata  &  Young~\&~Lo~(1996) \\      
NGC 3079 &     0.29$\,\times\,$0.25 &    9.7$\,\times\,$2.0   &   9.7 &   1125 &  16.4 &  9.86 &  SB(s)c          &\nodata  &  Irwin~\&~Seaquist~(1991) \\
NGC 3198 &     0.58$\,\times\,$0.42 &   21.0$\,\times\,$7.0   &  14.0 &    663 &   7.7 &  9.53 &  SB(rs)c         &\nodata  &  Begeman~(1989) \\          
IC 2574 &      0.50$\,\times\,$0.50 &   19.0$\,\times\,$10.0  &  15.0 &     57 &   3.0 &  8.78 &  SAB(s)m         &\nodata  &  Martinbeau~et~al.~(1994) \\
UGC 5716 &     0.54$\,\times\,$0.53 &    3.5$\,\times\,$2.3   &   3.1 &   1277 &  16.3 &  8.76 &  Sm:             &\nodata  &  van~Zee~et~al.~(1997) \\   
UGC 5764 &     0.53$\,\times\,$0.50 &    3.8$\,\times\,$1.8   &   3.0 &    586 &   5.0 &  7.79 &  IB(s)m:         &\nodata  &  van~Zee~et~al.~(1997) \\   
UGC 5829 &     0.33$\,\times\,$0.33 &    6.0$\,\times\,$5.3   & \nodata &    629 &   5.7 &  8.68 &  Im              &\nodata  &  Taylor~et~al.~(1994) \\    
KDG 73 &       0.20$\,\times\,$0.17 &    1.6$\,\times\,$0.5   & \nodata &    113 &   3.6 &  6.32 &  Im              & mult  &  Ott~et~al.~(2002) \\       
NGC 3510 &     1.03$\,\times\,$0.95 &    6.5$\,\times\,$2.0   & \nodata &    705 &   5.9 &  8.59 &  SB(s)m          &\nodata  &  Taylor~et~al.~(1994) \\    
NGC 3626 &     0.68$\,\times\,$0.22 &    8.0$\,\times\,$0.8   & \nodata &   1493 &  19.7 &  8.89 &  (R)SA(rs)0+     & mult  &  van~Driel~et~al.~(1989) \\ 
NGC 3628 &     0.50$\,\times\,$0.50 &   12.0$\,\times\,$2.1   &  10.1 &    843 &   6.2 &  9.38 &  SAb:            &\nodata  &  Wilding~et~al.~(1993) \\   
NGC 3726 &     0.63$\,\times\,$0.47 &   12.0$\,\times\,$5.0   &   7.8 &    866 &  11.6 &  9.54 &  SAB(r)c         &\nodata  &  Wevers~et~al.~(1986) \\    
UGC 6578 &     0.17$\,\times\,$0.14 &    1.3$\,\times\,$1.1   &   1.2 &   1099 &   9.7 &  7.92 &  pair            &\nodata  &  van~Zee~et~al.~(1998) \\   
NGC 3900 &     0.90$\,\times\,$0.42 &    5.2$\,\times\,$2.4   &   5.8 &   1798 &  28.5 &  9.50 &  SA(r)0+         &\nodata  &  van~Driel~et~al.~(1989) \\ 
UM 461 &       0.18$\,\times\,$0.16 &    2.4$\,\times\,$1.6   &   2.1 &   1039 &   7.9 &  7.96 &  BCD/Irr         &\nodata  &  van~Zee~et~al.~(1998) \\   
UGC 6850 &     0.17$\,\times\,$0.14 &    1.3$\,\times\,$1.2   &   1.5 &   1055 &   8.2 &  7.96 &  Pec             &\nodata  &  van~Zee~et~al.~(1998) \\   
NGC 3941 &     0.67$\,\times\,$0.42 &    7.0$\,\times\,$1.5   & \nodata &    928 &  11.7 &  8.76 &  SB(s)$0^0$      & ring  &  van~Driel~\&~van~Woerden~(1989) \\
M 109 &        0.44$\,\times\,$0.33 &   10.5$\,\times\,$5.0   &   8.7 &   1048 &  15.7 &  9.67 &  SB(rs)bc        &\nodata  &  Gottesman~et~al.~(1984) \\ 
NGC 4151 &     0.36$\,\times\,$0.29 &    7.5$\,\times\,$7.5   &   9.7 &    995 &  13.5 &  9.53 &  (R')SAB(rs)ab:  &\nodata  &  Pedlar~et~al.~(1992) \\    
UGC 7178 &     0.57$\,\times\,$0.53 &    3.2$\,\times\,$3.0   &   3.2 &   1339 &  16.0 &  8.70 &  IAB(rs)m:       &\nodata  &  van~Zee~et~al.~(1997) \\   
M 98 &         1.17$\,\times\,$0.80 &   13.5$\,\times\,$3.0   &   7.0 &   -142 &  17.0 &  9.72 &  SAB(s)ab        &\nodata  &  Cayatte~et~al.~(1990) \\   
NGC 4216 &     0.80$\,\times\,$0.78 &    7.0$\,\times\,$1.7   &   5.5 &    131 &  17.0 &  9.35 &  SAB(s)b:        &\nodata  &  Cayatte~et~al.~(1990) \\   
NGC 4222 &     0.80$\,\times\,$0.78 &    4.5$\,\times\,$1.5   & \nodata &    230 &  17.0 &  8.95 &  Sc              &\nodata  &  Cayatte~et~al.~(1990) \\   
NGC 4242 &     0.58$\,\times\,$0.42 &    6.0$\,\times\,$4.0   &   5.5 &    517 &   5.6 &  8.44 &  SAB(s)dm        &\nodata  &  Wevers~et~al.~(1986) \\    
M 99 &         1.08$\,\times\,$0.70 &    7.5$\,\times\,$5.0   &   6.3 &   2407 &  34.6 & 10.33 &  SA(s)c          &\nodata  &  Cayatte~et~al.~(1990) \\   
M 61 &         0.75$\,\times\,$0.73 &    8.5$\,\times\,$6.0   &   7.3 &   1566 &  19.5 &  9.87 &  SAB(rs)bc       &\nodata  &  Cayatte~et~al.~(1990) \\   
M 100 &        0.75$\,\times\,$0.70 &    6.5$\,\times\,$5.5   &   6.0 &   1571 &  20.4 &  9.68 &  SAB(s)bc        &\nodata  &  Cayatte~et~al.~(1990) \\   
NGC 4395 &     0.80$\,\times\,$0.43 &   16.0$\,\times\,$11.0  &  15.2 &    319 &   3.0 &  8.81 &  SA(s)m:         &\nodata  &  Wevers~et~al.~(1986) \\    
NGC 4402 &     0.35$\,\times\,$0.28 &    2.1$\,\times\,$0.6   &   1.8 &    232 &  17.0 &  8.68 &  Sb              &\nodata  &  Cayatte~et~al.~(1990) \\   
M 88 &         0.67$\,\times\,$0.65 &    6.5$\,\times\,$2.5   &   4.7 &   2281 &  33.4 &  9.82 &  SA(rs)b         &\nodata  &  Cayatte~et~al.~(1990) \\   
NGC 4535 &     0.70$\,\times\,$0.68 &    7.0$\,\times\,$4.5   &   6.8 &   1961 &  27.9 & 10.11 &  SAB(s)c         &\nodata  &  Cayatte~et~al.~(1990) \\   
IC 3522 &      0.40$\,\times\,$0.40 &    3.0$\,\times\,$2.1   & \nodata &    668 &   8.1 &  8.19 &  ImIII-IV        &\nodata  &  Skillman~et~al.~(1987) \\  
NGC 4568 &     0.67$\,\times\,$0.67 &    4.0$\,\times\,$2.5   & \nodata &   2255 &  33.5 &  9.50 &  SA(rs)bc        &\nodata  &  Cayatte~et~al.~(1990) \\   
M 58 &         0.32$\,\times\,$0.30 &    3.5$\,\times\,$1.0   &   4.0 &   1519 &  19.6 &  8.88 &  SAB(rs)b        & mult  &  Cayatte~et~al.~(1990) \\   
NGC 4654 &     0.80$\,\times\,$0.72 &    5.5$\,\times\,$3.5   &   5.3 &   1037 &  13.3 &  9.33 &  SAB(rs)cd       &\nodata  &  Cayatte~et~al.~(1990) \\   
UGC 7906 &     0.40$\,\times\,$0.40 &    2.4$\,\times\,$1.1   & \nodata &   1010 &  12.7 &  8.19 &  ImIV            &\nodata  &  Skillman~et~al.~(1987) \\  
NGC 4725 &     0.97$\,\times\,$0.42 &   12.0$\,\times\,$7.0   &  11.0 &   1206 &  16.7 &  9.85 &  SAB(r)ab        &\nodata  &  Wevers~et~al.~(1986) \\    
M 94 &         0.83$\,\times\,$0.83 &    4.6$\,\times\,$3.3   &   5.7 &    308 &   3.5 &  8.43 &  (R)SA(r)ab      &\nodata  &  Mulder~\&~van~Driel~(1993) \\
NGC 4731 &     0.55$\,\times\,$0.38 &    8.5$\,\times\,$6.0   & \nodata &   1495 &  17.3 &  9.87 &  SB(s)cd         &\nodata  &  Gottesman~et~al.~(1984) \\ 
NGC 4789A &    0.75$\,\times\,$0.75 &   10.0$\,\times\,$4.0   &   7.2 &    374 &   3.1 &  8.21 &  IB(s)mIV-V      &\nodata  &  Carignan~\&~Beaulieu~(1989) \\
UGC 8091 &     0.29$\,\times\,$0.29 &    2.2$\,\times\,$2.1   & \nodata &    214 &   2.2 &  6.99 &  ImV             &\nodata  &  Lo~et~al.~(1993) \\        
NGC 5033 &     0.72$\,\times\,$0.43 &   13.0$\,\times\,$6.0   &  11.8 &    875 &  11.4 &  9.83 &  SA(s)c          &\nodata  &  Wevers~et~al.~(1986) \\    
M 63 &         1.22$\,\times\,$0.82 &   30.0$\,\times\,$13.0  &  24.7 &    504 &   5.4 &  9.57 &  SA(rs)bc        &\nodata  &  Bosma~(1981) \\            
UGC 8333 &     0.40$\,\times\,$0.33 &    4.6$\,\times\,$1.4   & \nodata &    935 &  11.3 &  8.60 &  Im:             &\nodata  &  Lake~et~al.~(1990) \\      
NGC 5101 &     0.67$\,\times\,$0.67 &    6.5$\,\times\,$1.5   &   6.8 &   1861 &  20.3 &  9.48 &  (R')SB(rl)0/a   & ring  &  van~Driel~et~al.~(1988b) \\
NGC 5102 &     0.62$\,\times\,$0.57 &   11.0$\,\times\,$4.5   & \nodata &    467 &   2.8 &  7.97 &  SA0-            &\nodata  &  van~Woerden~et~al.~(1993) \\
NGC 5371 &     0.72$\,\times\,$0.47 &    6.0$\,\times\,$3.0   &   4.3 &   2553 &  38.9 &  9.97 &  SAB(rs)bc       &\nodata  &  Wevers~et~al.~(1986) \\    
NGC 5585 &     0.52$\,\times\,$0.52 &   11.6$\,\times\,$6.4   &   8.6 &    305 &   5.0 &  8.98 &  SAB(s)d         &\nodata  &  Cote~et~al.~(1991) \\      
NGC 6503 &     0.43$\,\times\,$0.42 &   22.0$\,\times\,$6.0   &  13.2 &     60 &   3.8 &  8.82 &  SA(s)cd         &\nodata  &  Wevers~et~al.~(1986) \\    
ESO 594-G004 & 0.37$\,\times\,$0.25 &    4.1$\,\times\,$2.2   & \nodata &    -77 &   1.1 &  6.94 &  IB(s)m:         & mult  &  Young~\&~Lo~(1997) \\      
NGC 6814 &     0.86$\,\times\,$0.70 &    5.3$\,\times\,$5.2   &   5.7 &   1563 &  20.6 &  9.47 &  SAB(rs)bc       &\nodata  &  Liszt~\&~Dickey~(1995) \\  
NGC 6946 &     0.67$\,\times\,$0.62 &   23.0$\,\times\,$17.0  &  20.0 &     48 &   7.0 & 10.01 &  SAB(rs)cd       &\nodata  &  Tacconi~\&~Young~(1986) \\ 
UGC 11820 &    0.73$\,\times\,$0.59 &    4.9$\,\times\,$2.9   &   5.2 &   1104 &  18.7 &  9.50 &  Sm              &\nodata  &  van~Zee~et~al.~(1997) \\   
NGC 7331 &     0.75$\,\times\,$0.42 &   15.0$\,\times\,$3.5   &  13.6 &    816 &  15.8 & 10.02 &  SA(s)b          &\nodata  &  Bosma~(1981) \\            
IC 5267 &      0.67$\,\times\,$0.67 &   13.0$\,\times\,$1.4   &   1.7 &   1713 &  27.5 &  9.63 &  (R)SA(rs)0/a    & ring  &  van~Driel~et~al.~(1988b) \\
UGC 12613 &    0.24$\,\times\,$0.24 &    4.5$\,\times\,$1.8   & \nodata &   -183 &   0.8 &  6.42 &  dIrr/dSph       &\nodata  &  Lo~et~al.~(1993) \\        
NGC 7793 &     0.75$\,\times\,$0.75 &   13.0$\,\times\,$8.0   &  13.2 &    230 &   3.4 &  8.88 &  SA(s)d          &\nodata  &  Carignan~\&~Puche~(1990a) \\
\enddata
\end{deluxetable}

\end{document}